\documentclass[a4paper,11pt]{article}

\usepackage{jinstpub} 
\usepackage{hyperref}
\usepackage{caption}
\usepackage{subcaption}
\usepackage{lineno}

\title{\boldmath Towards a large-area RPWELL detector: design optimization and performance}

\author[a,b,1]{D. Zavazieva,\note{Corresponding author.}}
\author[b]{L. Moleri,}
\author[b]{A. Jash,}
\author[b]{G. Sela,}
\author[b]{F. de Vito-Halevy}
\author[b]{and S. Bressler}

\affiliation[a]{Department of Nuclear Engineering, Ben-Gurion University of the Negev,\\Be'er Sheva, Israel}
\affiliation[b]{Department of Particle Physics and Astrophysics, Weizmann Institute of Science,\\Rehovot, Israel}

\emailAdd{darinaza@post.bgu.ac.il}

\abstract{We present a new design and assembly procedure of a large-area gas-avalanche Resistive-Plate WELL (RPWELL) detector. A $50\times50 ~\mathrm{cm^2}$ prototype was tested in $\mathrm{80 ~GeV/c}$ muon beam at CERN-SPS, presenting improved performances compared to previous ones: MIP detection efficiency over 96\%  with 3\% uniformity across the entire detector area, a charge gain of $\mathrm{\approx{7.5 \times 10^3}}$ with a uniformity of 22\%, and discharge probability below $\mathrm{10^{-6}}$ with a few single hotspots attributed to production imperfections. These results pave the way towards further up-scaling detectors of this kind.}


\arxivnumber{} 

\begin{document}
\maketitle
\flushbottom

\section{Introduction}
\label{sec:intro}

Being cost-effective, gaseous radiation detectors are suitable for applications requiring large area coverage. 
However, scaling up the dimensions of detector elements can be challenging. Large area detector design has to comply with the desired physics performance as well as with technical aspects like material budget, cost, and construction feasibility.

Micro-Pattern Gaseous Detectors (MPGDs) are potentially easy to scale up: their construction strongly relies on common industrial technologies, providing flexibility and reliability of the processes. Recent examples are the large area triple GEM \cite{Bianco:2020bem} and Micromegas chambers \cite{Gnesi:2020umd} that were successfully deployed and commissioned in the CMS and ATLAS muon systems at CERN, respectively.

The Thick Gaseous Electron Multiplier (THGEM) \cite{Chechik:2004wq} technology and its derivatives are attractive for various applications requiring cost-effective solutions for large-area coverage; they provide high detection efficiency at irradiation flux up to $\mathrm{10^4 ~Hz/cm^2}$ \cite{Moleri:2016bjv} at moderate energy resolution (of the order of 20\% FWHM, measured with a 5.9 keV mono-energetic X-ray source), position resolution (at the level of $300 ~\mathrm{\mu m}$ \cite{Moleri:2017qhi}) and time resolution (about 10 ns for MIPs \cite{Alon:2008pq}). For a recent review on THGEM detectors see \cite{Bressler:2023wrl}. 

\begin{figure}[htbp]

\includegraphics[width=.85\textwidth]{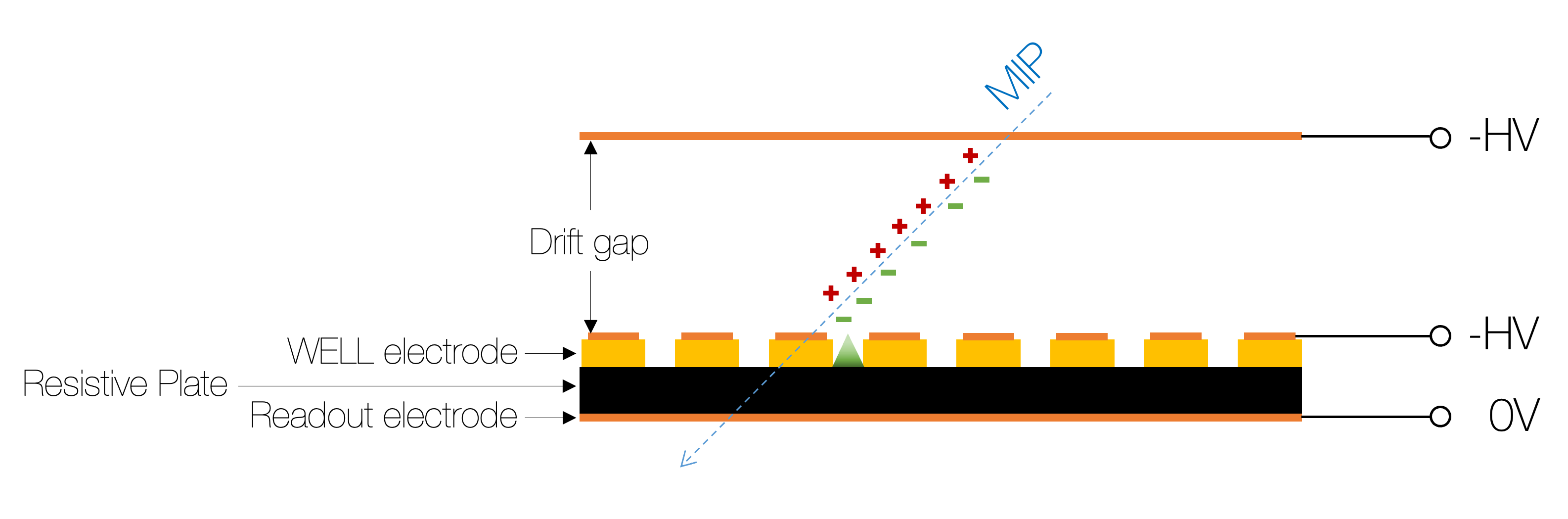}
\caption{\label{fig:a} RPWELL detector schematics. Radiation-induced primary electrons drift along the filed lines into the THWELL holes where they undergo charge-avalanche multiplication. Signals are induced on the anode by the charges' movement \protect{\cite{Shockley:1938itm, Ramo:1939vr}}.}
\end{figure}

The THWELL electrode is a THGEM Cu-clad on one side only \cite{Arazi:2013hdn}. In a THWELL detector configuration, the single-sided electrode is coupled directly to the readout anode. The Resistive WELL (RWELL) \cite{Arazi:2013hdn} and the Resistive Plate WELL (RPWELL) \cite{Rubin:2013jna} are THWELL-like structures with resistive elements embedded in the detector design to mitigate the effect of discharges \cite{Jash:2022bxy}. In the RWELL, the THWELL-electrode is coupled to the readout anode through a resistive layer $\mathrm{(\sim100K - 10M ~\Omega/\square)}$ deposited on a thin $\mathrm{(\sim 100 ~ \mu m)}$ insulating sheet. The RPWELL detector discussed in this work is shown schematically in Figure \ref{fig:a}. In this configuration, the THWELL electrode is coupled to the readout anode through a material of high bulk resistivity ($\mathrm{\sim 10^9 \text{-} 10^{12} ~\Omega cm}$).

Similar to other gaseous-based technologies, scaling up the dimensions of THGEM-based detectors poses the challenges of maintaining uniform response over the entire detector area and ensuring electrical stability. The latter is easier to achieve with resistive configurations, RWELL and RPWELL. However, the resistive configuration also poses technical challenges --- e.g., the need to attach the single-sided THGEM electrode to the resistive material without jeopardizing the holes' quality. So far, large area THWELL- and THGEM-based detectors were studied for Digital Hadron Calorimetry (DHCAL) \cite{f, Shaked-Renous:2022kxo, Bressler:2019uyt} and ring Cherenkov imaging \cite{Agarwala:2022xpi}, respectively.

RWELL detector with the resistive layer made of Diamond-Like Carbon, DLC,  was studied in Ref. \cite{Hong:2020hwx, f}. Prototypes of  $25\times25 ~\mathrm{cm^2}$ area were constructed and characterized in terms of gain and rate capability. Twenty sectors of $5 \times 25 ~\mathrm{cm^2}$ area were printed on single boards to construct RWELL prototypes of $50\times100 ~\mathrm{cm^2}$ \cite{f}. The transition region in between the sectors was coated with insulating glue on both the tiles and the anode to avoid a direct path between nearby conductors known to be a source of electrical instabilities. Preliminary results reported gain uniformity of about 15\%, rate capability up to $100 ~\mathrm{kHz/cm^2}$, and Minimum Ionizing Particle (MIP) detection efficiency greater than 95\%.

In recent years, an effort has been made to scale up the area of RPWELL detectors from $10 \times 10 ~\mathrm{cm^2}$ \cite{Bressler:2015dya} to $30 \times 30 ~\mathrm{cm^2}$ \cite{Moleri:2016hgk} and up to $50 \times 50 ~\mathrm{cm^2}$ \cite{Shaked-Renous:2022kxo}. The $30\times30 ~\mathrm{cm^2}$ RPWELL prototype was equipped with a polymer-made resistive plate\footnote{Semitron ESD225}, and the WELL electrode was mechanically pressed to the resistive plate using plastic pillars. This technique resulted in relatively large dead areas (5\%) and, in addition, some electrically weak spots were identified in the proximity of the pillars \cite{Moleri:2016hgk}. Four $\mathrm{25\time 25 ~cm^2}$ Fe-doped glass tiles, resistive plates
\cite{Wang:2010bg}, were used for the assembly of $50\times50 ~\mathrm{cm^2}$ prototypes \cite{Shaked-Renous:2022kxo}. The large, $50\times50 ~\mathrm{cm^2}$, WELL electrode used for their construction had large thickness variations (20\%) resulting in large gain non-uniformity (50\%). Furthermore, some of the glue used to attach the WELL electrode to the glass tiles penetrated the holes and induced electrical instabilities \cite{Bressler:2019uyt}.

In Section \ref{sec:design}, we present an improved design and assembly procedure of a $50\times50 ~\mathrm{cm^2}$ RPWELL detector which overcomes past limitations. A new detector prototype was built and tested with relativistic muons at the CERN SPS beam line. The testing procedure is detailed in Section \ref{sec:method}, followed by a report of the prototype's performance in Section \ref{sec:performance} and a discussion in Section \ref{sec:discussion}.

\section{Detector design \& Assembly}
\label{sec:design}

The new RPWELL design presented in this work addresses the two main weaknesses identified in the design and assembly of the prototypes used in \cite{Bressler:2019uyt,kk}: a) large electrode thickness non-uniformity resulting in large gain non-uniformities,  and b) penetration of glue into the THWELL holes becoming a source for discharges. Stringent Quality Assurance and Quality Control (QA\&QC) requirements are added in the electrode selection process --- $+0/-5 \%$ of the total electrode thickness is enforced. A new construction approach where the glue is applied only in dedicated spots is developed to prevent the glue from spreading into the THWELL holes.   

\begin{figure}[htbp]
\centering
\includegraphics[width=0.9\textwidth]{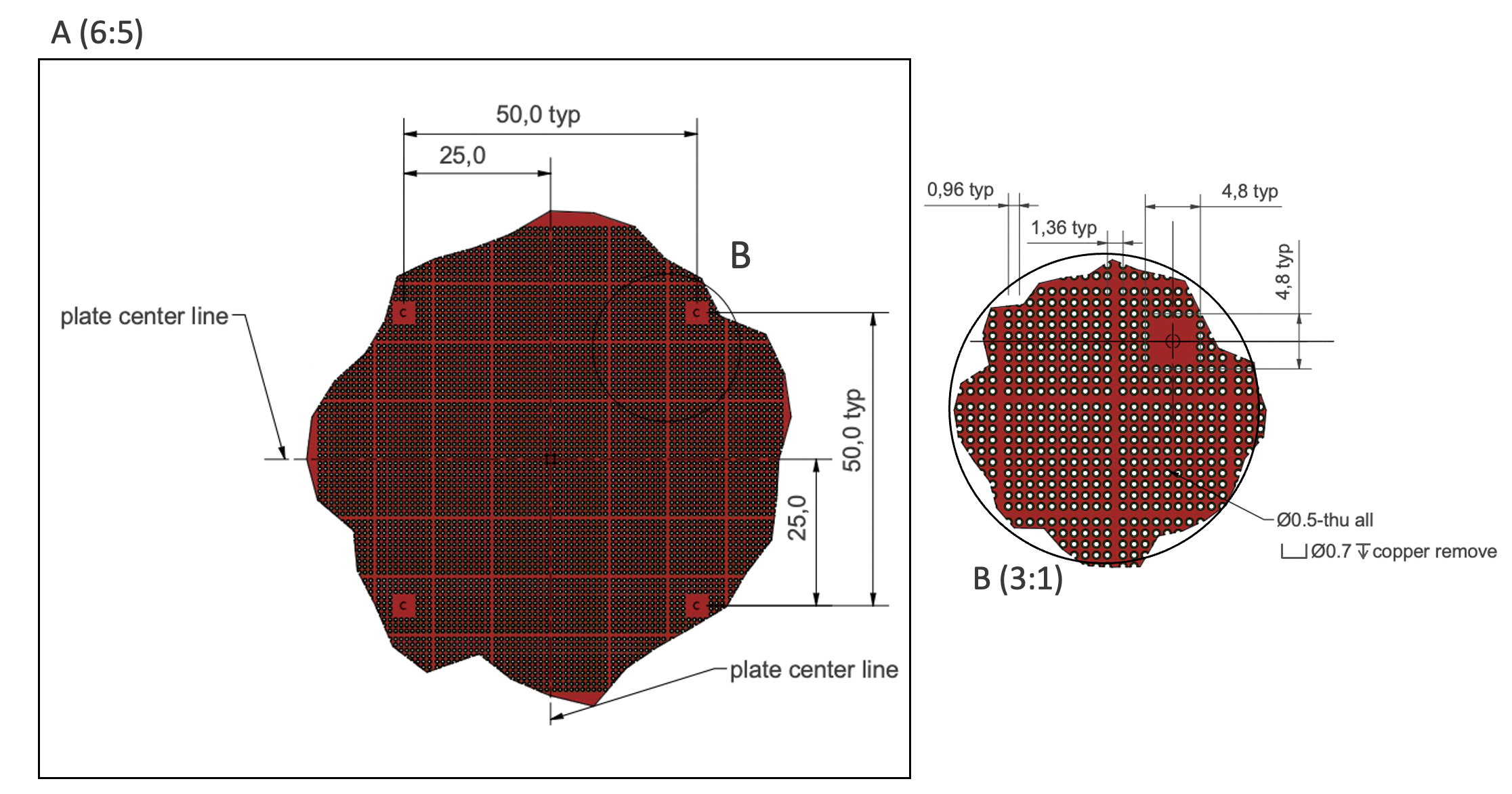}
\caption{\label{fig:b} CAD drawing of THGEM board. A – pattern of gluing points across the area, B – dimensions of the gluing point.}
\end{figure}

Following the new design, a $50\times50 ~\mathrm{cm^2}$ RPWELL prototype was assembled. Four $\mathrm{\mathrm{25 \times 25}~cm^2}$ Fe-doped glass tiles (bulk resistivity of $\mathrm{2\times 10^9 ~\Omega cm}$) were used to comprise the resistive plate. The THWELL electrode was produced by Eltos\footnote{Eltos S.p.A., Strada E 44, San Zeno 52100 Arezzo, Italy} with cleaning post-treatment at the CERN micro-pattern technologies laboratory \cite{Bressler:2023wrl}. In view of potential usage as DHCAL sampling elements \cite{Shaked-Renous:2022kxo, f, Bressler:2019uyt}, $0.4 ~\mathrm{mm}$ thick electrode perforated with $0.5 ~\mathrm{mm}$ diameter holes and $0.1 ~\mathrm{mm}$ rim etched around them was used. The holes were arranged in a square lattice such that each array of $10\times10$ holes creates a $1\times1 ~\mathrm{cm^2}$ pad region. The pads are separated by a Cu grid. Areas equivalent to $4\times4$ holes in the middle of a pad region every $5 ~\mathrm{cm}$ were left plain to serve as gluing spots of the electrode onto the resistive plate (Figure \ref{fig:b}). The spread of the glue in the proximity of the gluing points depends on the amount of glue and the pressure applied during the curing. An accurate procedure is needed to ensure a well-controlled spread.

The assembly procedure is detailed in Appendix \ref{Appendix A}. It consists of the following steps:
\begin{enumerate}
    \item Gluing the four glass tiles to the readout plane using an epoxy-graphite mixture to ensure electrical contact between the two for efficient charge evacuation;
    \item Protecting the interface between the glass tiles with insulating paste to avoid a direct path between the THWELL-top and the anode leading to discharges;
    \item Gluing the THWELL bottom to the glass in the dedicated points and curing under vacuum;
    \item Fixing $3 ~\mathrm{mm}$ thick drift gap by gluing side frames;
    \item Closing the chamber with an FR4 plate with an internal copper-clad part serving as a cathode plane;
    \item Sealing the chamber to ensure gas-tight volume. 
\end{enumerate}
The readout anode was segmented in $1 ~\mathrm{mm}$ pitch 1D strips routed to 4 multi-pin connectors to record the signals from the strips.

\section{Experimental setup \& Methodology}
\label{sec:method}

\begin{figure}[htbp]
\centering
\includegraphics[width=.4\textwidth]{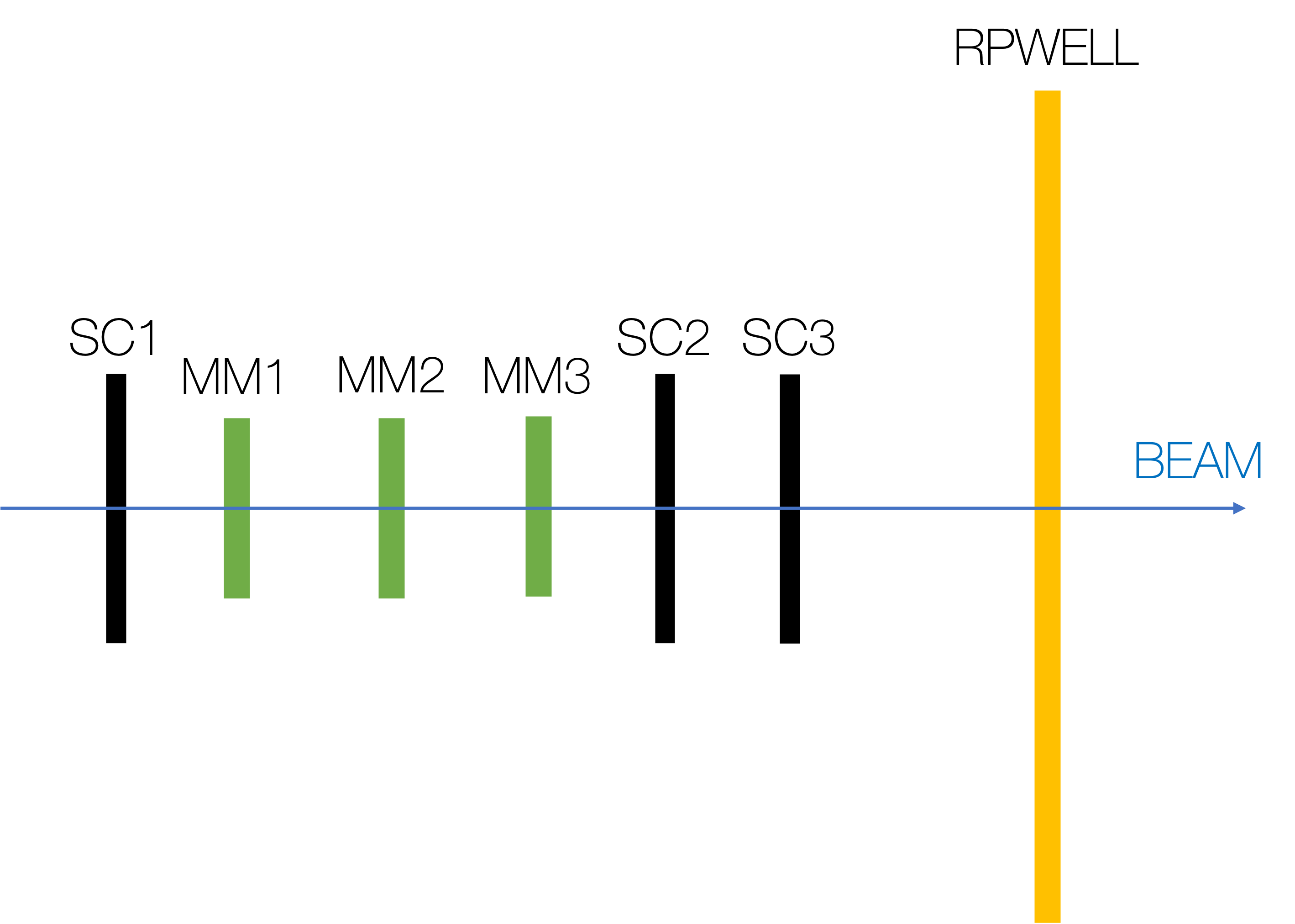}
\qquad
\includegraphics[width=.47\textwidth]{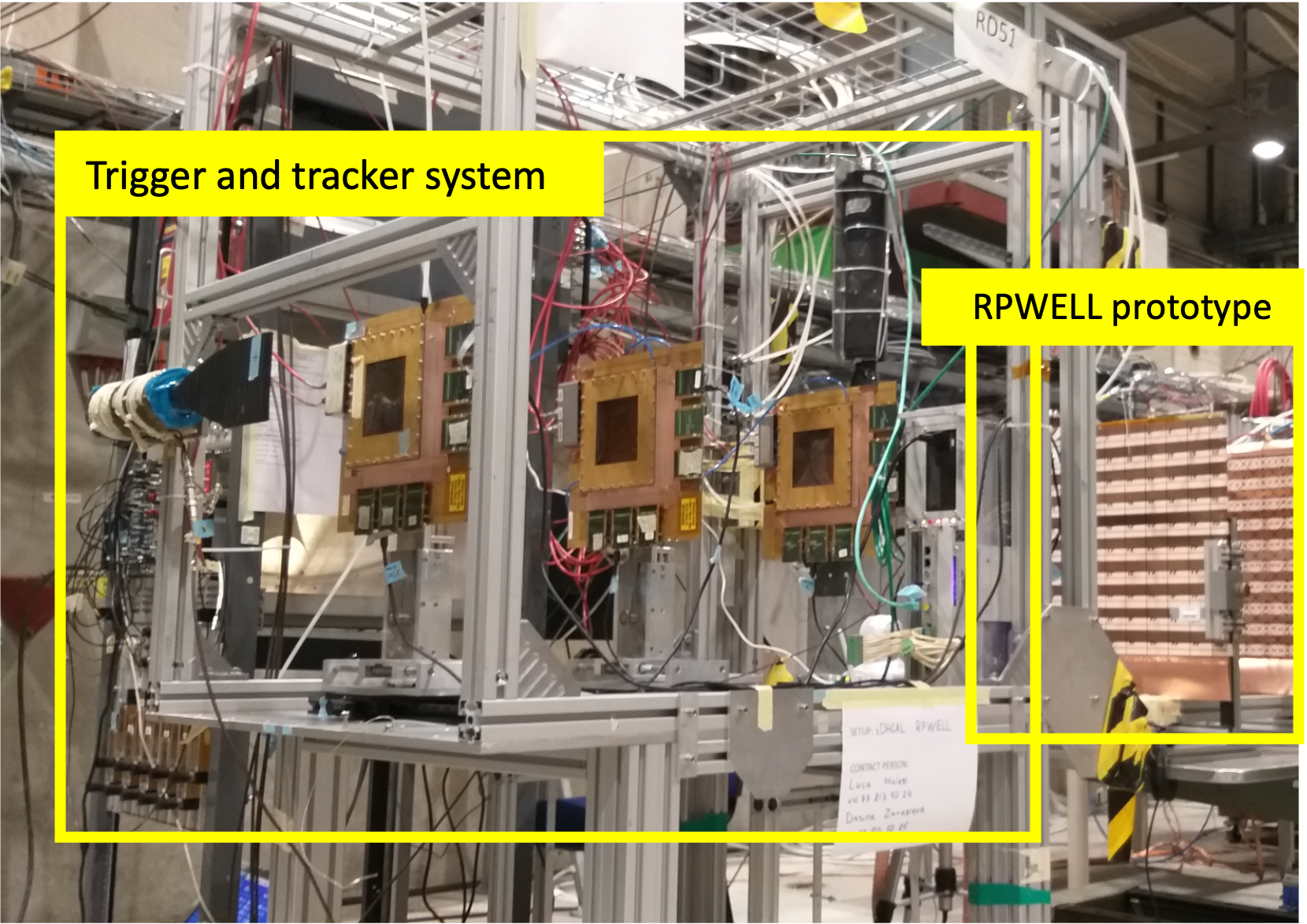}
\caption{\label{fig:setup} Schematics of the experimental setup. SC – scintillator, MM – Micromegas detector, RPWELL – RPWELL detector under test. }
\end{figure}

Measurements were carried out with $\mathrm{80 ~GeV/c}$ muons at the CERN-SPS north area H4 beam line. The beam covered an area of about $\mathrm{6\times6 ~cm^2}$ as measured with dedicated tracking detectors. The experimental setup was similar to the one employed in \cite{Moleri:2016hgk}. It is shown schematically in Figure {\ref{fig:setup}}.

The tracking system comprised three Micromegas chambers of $\mathrm{6\times6 ~cm^2}$ active area with 2D readout strips of $250 ~\mu\mathrm{m}$ pitch. A reference trigger signal was obtained by three scintillators arranged in coincidence to define a $\mathrm{6\times6 ~cm^2}$ area. The trigger rate was of the order of $\mathrm{2\times10^4 ~muons/spill}$ with a spill duration of $\mathrm{4-5 ~seconds}$. The RPWELL prototype was fixed on an X-Y movable table, and its entire area was scanned. At each measurement point about $3 \times 10^4$ events were recorded.

\subsection{RPWELL operation}
The detector was operated with a pre-mixed $\mathrm{Ar/CO_2}$ (93:7) gas mixture at a flow of $\mathrm{5 ~l/h}$. High voltage was supplied to the drift and THWELL-top electrode by a CAEN power supply (A1821N). The drift gap of $3 ~\mathrm{mm}$ was kept at a constant field value of $\mathrm{0.5 ~kV/cm}$ while the RPWELL was operated at 1350 V, corresponding to an amplification field (parallel plate equivalent) of $\mathrm{33.75 ~kV/cm}$ for most of the measurements.

\subsection{Readout system}

The signals from both the tracker and the RPWELL detector were recorded using APV25 ASIC chips \cite{Martoiu:2011zja} and read out using the Scalable Readout System (SRS) \cite{Martoiu:2013aca}. The data acquisition was triggered by the scintillator reference signals at a reduced rate of $\mathrm{700-800 ~counts/spill}$ due to a limitation of the data acquisition system. Twelve bits ADC was used to digitize each triggered event, and the pulses were sliced in 15 bins of $\mathrm{25 ~ns}$. Only signals passing a Zero Suppression threshold as detailed in \cite{Bressler:2015dya} were stored.

The APV25 chip has a shaping time of $\mathrm{50 ~ns}$. As shown in \cite{Arazi:2013hdn, Rubin:2013jna}, the rise-time of the RPWELL detector is $1 - 2 ~\mathrm{\mu s}$. It comprises a fast component (up to $\mathrm{100 ~ns}$ \cite{Bhattacharya:2018sqx}) and a slow one arising from the motion of avalanche ions inside the hole; the latter carries about 80\% of the total amplitude. Consequently, with $\mathrm{50 ~ns}$ shaping time, about 10\% of the RPWELL signal is integrated, and results are reported in terms of the effective gain which is estimated to be about an order of magnitude lower than the gas gain of the detector.

\subsection{Analysis framework}

The MM and the RPWELL signals were acquired with the mmdaq software package \cite{mmdaq} as detailed in \cite{Bressler:2015dya}. For both technologies, hits from neighboring strips were grouped into clusters. The cluster position was defined as the charge-weighted average strip position. The RPWELL ADC counts were converted to charge following the calibration procedure detailed in \cite{Moleri:2016hgk}. The cluster charge was defined as the sum of all charges measured from all the strips in a cluster. Using the software package described in \cite{hltt}, MM clusters were used to form a track. The tracks were extrapolated to define their expected intersect with the RPWELL plane. An RPWELL cluster was defined as matching to a track if both were triggered by the same trigger, and if the distance between the RPWELL track and the intersection point was smaller than 10 mm. 

\subsubsection{Threshold optimization}

Dedicated pedestal runs, with no beam, were taken in between physics measurements. The data collected in these runs were used to define the baseline and the noise level of each APV25 channel; the former was derived as the mean value of the signals measured in each channel, and the latter as the standard deviation, $\mathrm{\sigma}$, of this distribution. A common Zero Suppression Factor (ZSF) relative to the noise level was used to set a threshold for all readout channels such that signals were stored only if the integral of the signal was greater than $\mathrm{ZSF\times\sigma\times n_{bins}}$, where $\mathrm{n_{bins}}$ is a number of time bins.

\begin{figure}
\centering
\begin{subfigure}[b]{0.435\textwidth}
\centering
\includegraphics[width=\textwidth]{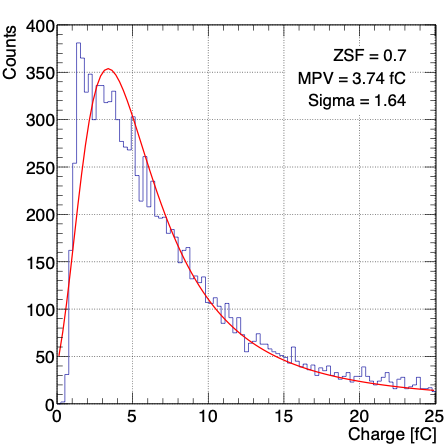}
\caption{}
\label{fig:i-a}
\end{subfigure}
\begin{subfigure}[b]{0.435\textwidth}
\centering
\includegraphics[width=\textwidth]{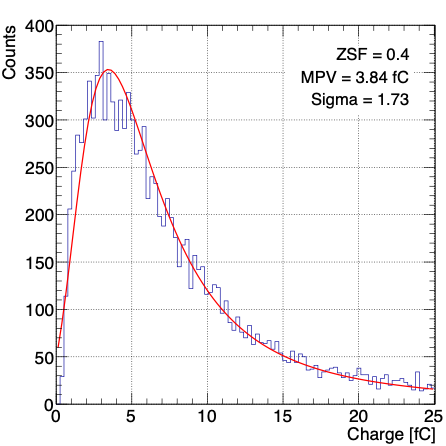}
\caption{}
\label{fig:i-b}
\end{subfigure}
\caption{\label{fig:i} Charge distributions for data processed with ZSF = 0.7 (a) and with ZSF = 0.4 (b). Red lines outline a Landau fit.}
\end{figure}

Following \cite{Bressler:2015dya, Moleri:2016hgk}, the data acquisition was performed with a low threshold of $\mathrm{ZSF = 0.4}$. After some optimization studies, the offline data processing was performed with the same ZSF value; a higher threshold, e.g., $\mathrm{ZSF = 0.7}$, aiming at reducing the noise level, resulted in the loss of low-charge signal hits, typically at the cluster edges. The latter led to a large fraction of clusters with low strip multiplicity and a pronounced mismeasurement of low-charge clusters. Consequently, as can be seen in Figure \ref{fig:i-a}, the charge spectra did not fit the Landau distribution\footnote{Only selected events, as explained in Subsection \ref{selection}, are shown.}.  At $\mathrm{ZSF = 0.4}$ such losses are negligible, and the Landau shape of the charge spectra is mostly restored, as seen in Figure \ref{fig:i-b}. The strip multiplicity increased by one in more than 43\% of the events and by two in more than 11\% of the events. Using lower ZSF, 0.4 relative to 0.7, an average charge of $0.6 ~\mathrm{fC}$ is gained.

\subsubsection{Event selection}
\label{selection}

\begin{figure}
\centering
\begin{subfigure}[b]{0.435\textwidth}
\centering
\includegraphics[width=\textwidth]{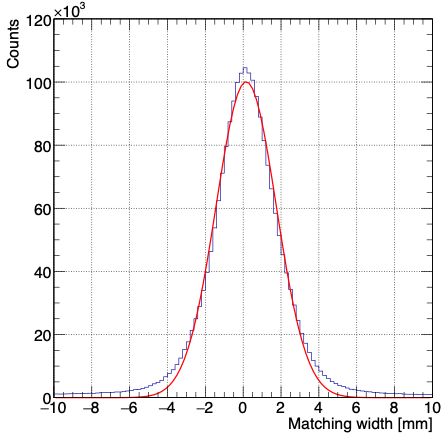}
\caption{}
\label{fig:k-a}
\end{subfigure}
\begin{subfigure}[b]{0.435\textwidth}
\centering
\includegraphics[width=\textwidth]{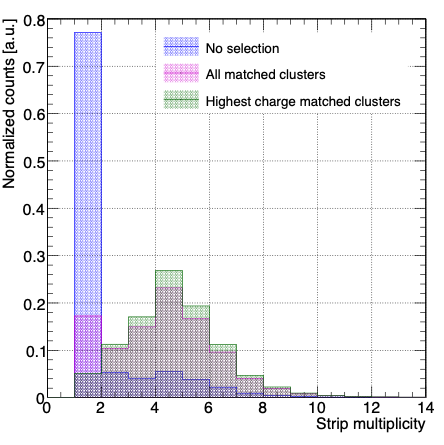}
\caption{}
\label{fig:k-b}
\end{subfigure}
\caption{\label{fig:k} Results of the event selection: a) distribution of residuals (distance between the selected cluster to an expected track position); b) change in the strip multiplicity with the applied steps of analysis (histograms are normalized to the total number of clusters).}
\end{figure}

An efficient muon event was defined as one in which a matching RPWELL cluster and a muon track were found. Since the RPWELL had a 1D readout, the matching could be done only in the x direction. The position of the RPWELL cluster in the y direction was set according to the position of the extrapolated track trajectory. If a track was matched to more than one RPWELL cluster, the cluster of higher charge was selected.  Figure \ref{fig:k-a} shows the distribution of the distance between the extrapolated track trajectory and the measured cluster position (i.e., residuals). The distribution fits well a Gaussian with a standard deviation of 1.6 mm. This indicates upon appropriate selection of events, but a spread wider than the RPWELL position resolution \cite{Moleri:2017qhi}. This is attributed to a misalignment of the RPWELL with respect to the tracker which does not affect the performance studies discussed in this work.

In Figure \ref{fig:k-b}, the strip multiplicity of the clusters is shown at different stages of the selection procedure. Before any selection the majority of clusters (77.1\%) are of strip multiplicity 1, while only 5.5\% are of strip multiplicity 4. The cluster-to-track matching requirement reduces the fraction of single-strip multiplicity clusters to 17.3\% and increases the fraction of four-strip multiplicity clusters to 23.2\%. Finally, when taking into account the cluster of higher charge in events with more than one matching cluster, only 5.1\% are of strip multiplicity 1, while 26.8\% are of strip multiplicity 4. 

\subsubsection{Noise estimation}

To estimate the noise level, several physics runs were acquired without the beam. The readout was triggered randomly at a fixed rate of $\mathrm{50 ~Hz}$. No clusters with >1 strip were found in these runs. The probability of 1 strip cluster occurrence was estimated at the level of 7\%. Since only 5\% of effective muon events are of strip multiplicity 1, the overall probability of noise hits to be counted in physics runs is at most at the level of $\mathrm{3.5 \times 10^{-3}}$, under the assumption that all the noise hits originate at the same point.

\section{Performance}
\label{sec:performance}

The detector performance is characterized in terms of three main parameters: detection efficiency, effective gain, and discharge probability. The detection efficiency was defined as the number of events in which an RPWELL cluster was matched to a track over the total number of triggered events. The effective gain was estimated from the cluster charge spectrum either as the average value of the distribution or as the Most Probable Value (MPV) of the Landau function best fitted to the spectrum. The discharge probability is defined as the fraction of a very high charge events saturating the APV channels occurring in triggered events and matched to the track. Within the resolution of the power supply monitoring (50 nA), these events did not lead to a voltage drop or current spikes, however, they increased significantly the average cluster charge.

\subsection{Efficiency}

\begin{figure}
\centering
\begin{subfigure}[b]{0.47\textwidth}
\centering
\includegraphics[width=\textwidth]{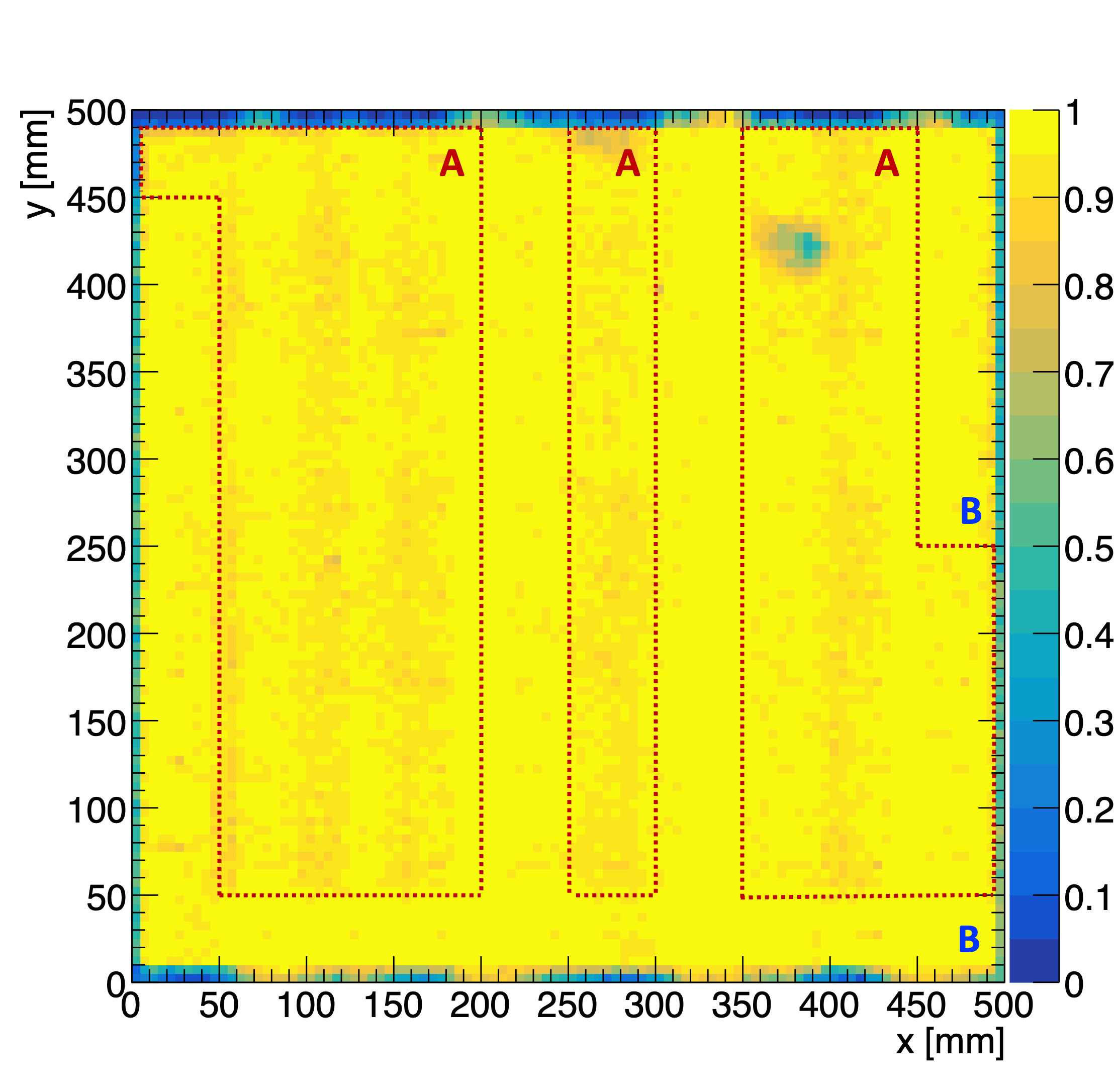}
\caption{}
\label{fig:l-a}
\end{subfigure}
\begin{subfigure}[b]{0.46\textwidth}
\centering
\includegraphics[width=\textwidth]{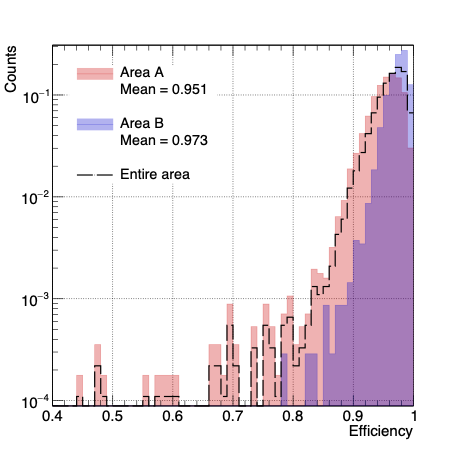}
\caption{}
\label{fig:l-b}
\end{subfigure}
\caption{\label{fig:l} Efficiency results: a) efficiency map; b) distribution of efficiency values for A and B areas marked on the map.}
\end{figure}

Figure \ref{fig:l-a} shows the efficiency map of the entire detector area. The individual values were averaged over an area of $\mathrm{5 \times 5 ~mm^2}$.  Overall, the efficiency is well above 90\% except for one circular area in the top right corner with a detached gluing spot. The Figure combines two sets of measurements. In the first, most of the detector area was scanned --- area A marked in the Figure. A repeated pattern of vertical lines of somewhat lower efficiency was identified. After improving the grounding, the measurements were repeated in some of the points --- area B. As expected, no low efficiency patterns are seen once a better signal-to-noise ratio is obtained. 

The distributions of the efficiency values are shown separately for the two sets of measurements in Figure \ref{fig:l-b}. In Area A where lower quality data is collected, the mean efficiency is 95.1\% with a spread of 3.6\%. In area B with higher quality data taken, the mean efficiency is 97.3\% with a spread of 1.8\%. For completeness, we report the overall efficiency to be 95.9\% with a variation of 3.2\%. 

\subsection{Gain}

\begin{figure}
\centering
\begin{subfigure}[b]{0.51\textwidth}
\centering
\includegraphics[width=\textwidth]{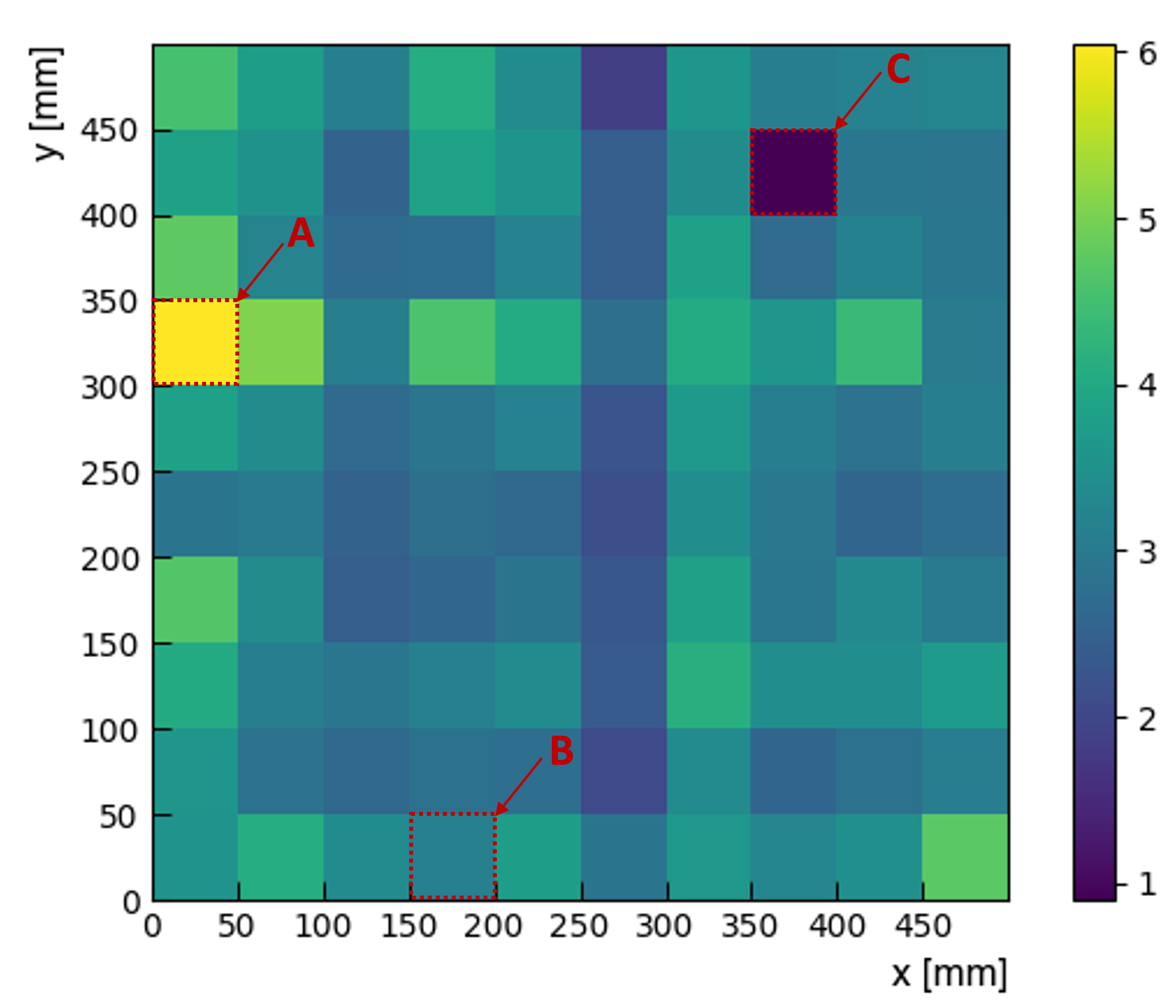}
\caption{}
\label{fig:m-a}
\end{subfigure}
\begin{subfigure}[b]{0.43\textwidth}
\centering
\includegraphics[width=\textwidth]{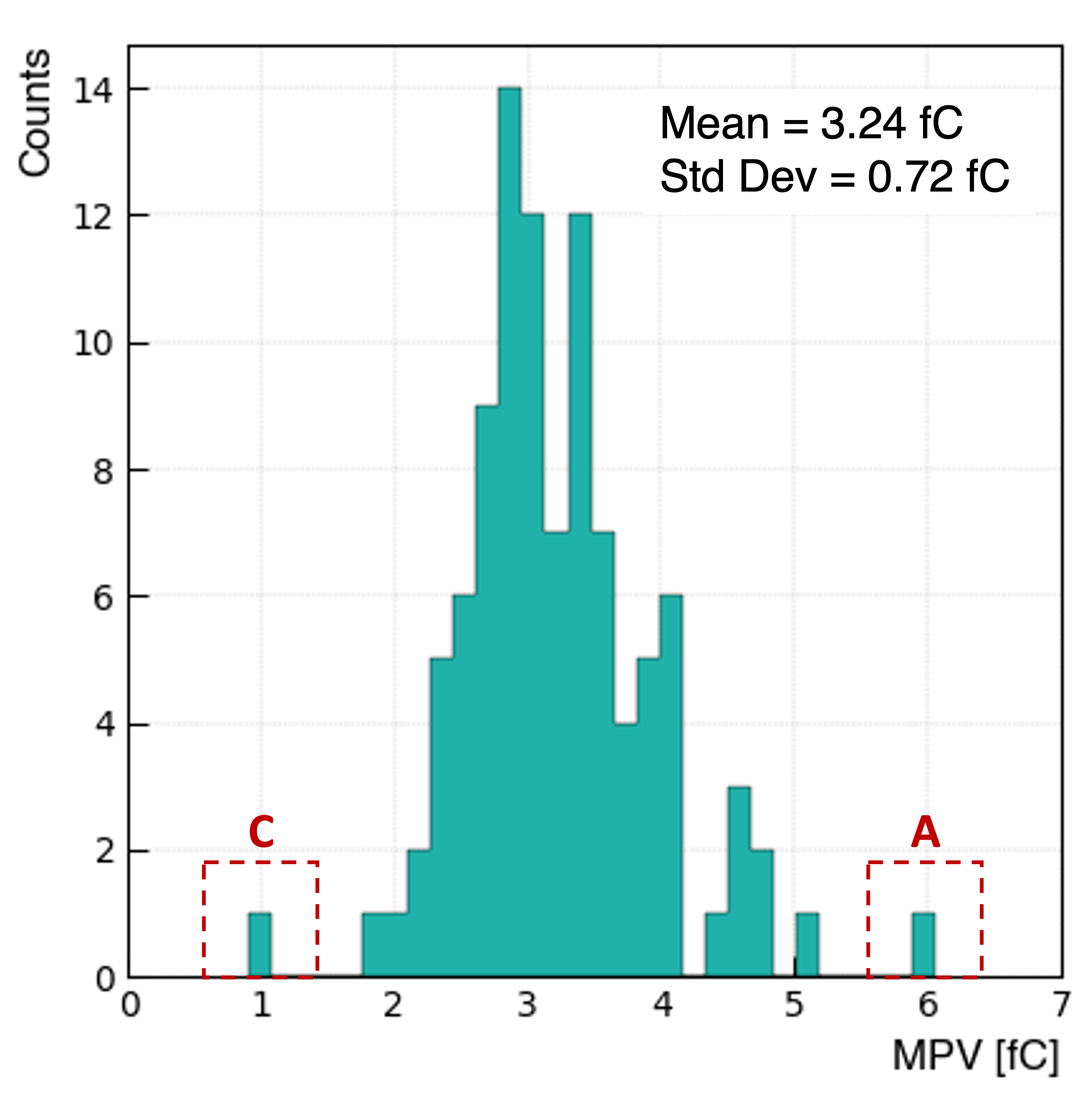}
\caption{}
\label{fig:m-b}
\end{subfigure}

\caption{\label{fig:m} Gain results: a) MPV map; b) distribution of MPV values.}
\end{figure}

To ensure sufficient statistics ($\mathrm{>10^4}$ events), the gain was evaluated from the charge spectrum measured over an area of each $5\times5 ~\mathrm{cm^2}$. The map of MPV values is shown in Figure \ref{fig:m-a}. Figure \ref{fig:m-b} shows their distribution; a mean of $3.24 ~\mathrm{fC}$ and the spread of 22\% are measured, the latter is consistent with the measured thickness variation ($\mathrm{+ 1.25 \% / \text{--} 5 \%} $). The lowest gain is measured in area C marked in Figure \ref{fig:m-a}, and corresponds to the position of the detached gluing point. The highest gain is measured in area A. 

\begin{figure}[htbp]
\centering
\includegraphics[width=.49\textwidth]{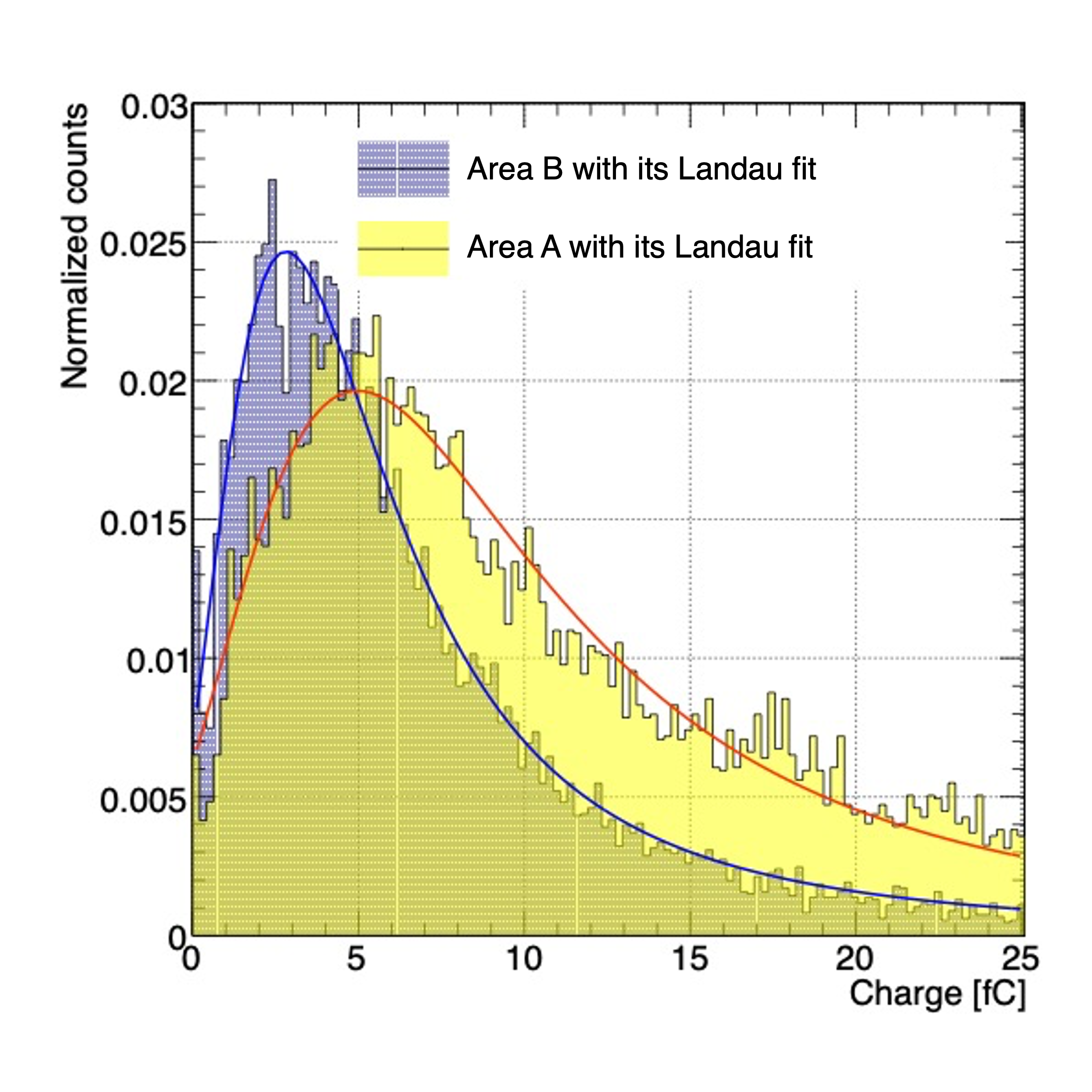}
\caption{\label{fig:n} Comparison of the charge distributions obtained at the lower and higher gain regions. Solid lines are Landau distribution fits.}
\end{figure}

An effective gain of $\approx 750$ is estimated assuming an average of 27 primary electrons. This number is estimated using HEED \cite{Smirnov:2005yi} where primary electrons were produced by 80 GeV muons traversing the detector perpendicular to its surface. As the fast shaping time of the APV25 ASIC allows it to collect only a part of the RPWELL signal, the gas gain is roughly estimated to be at the level of $7.5\times10^3$. As shown in Figure \ref{fig:n}, both high  ($6.05 ~\mathrm{fC}$ MPV) and low ($3.12~\mathrm{fC}$ MPV) charge spectra fit well Landau distribution, indicating upon normal and stable response of the detector. 

\subsection{Discharge probability}

\begin{figure}[htbp]
\centering
\includegraphics[width=.48\textwidth]{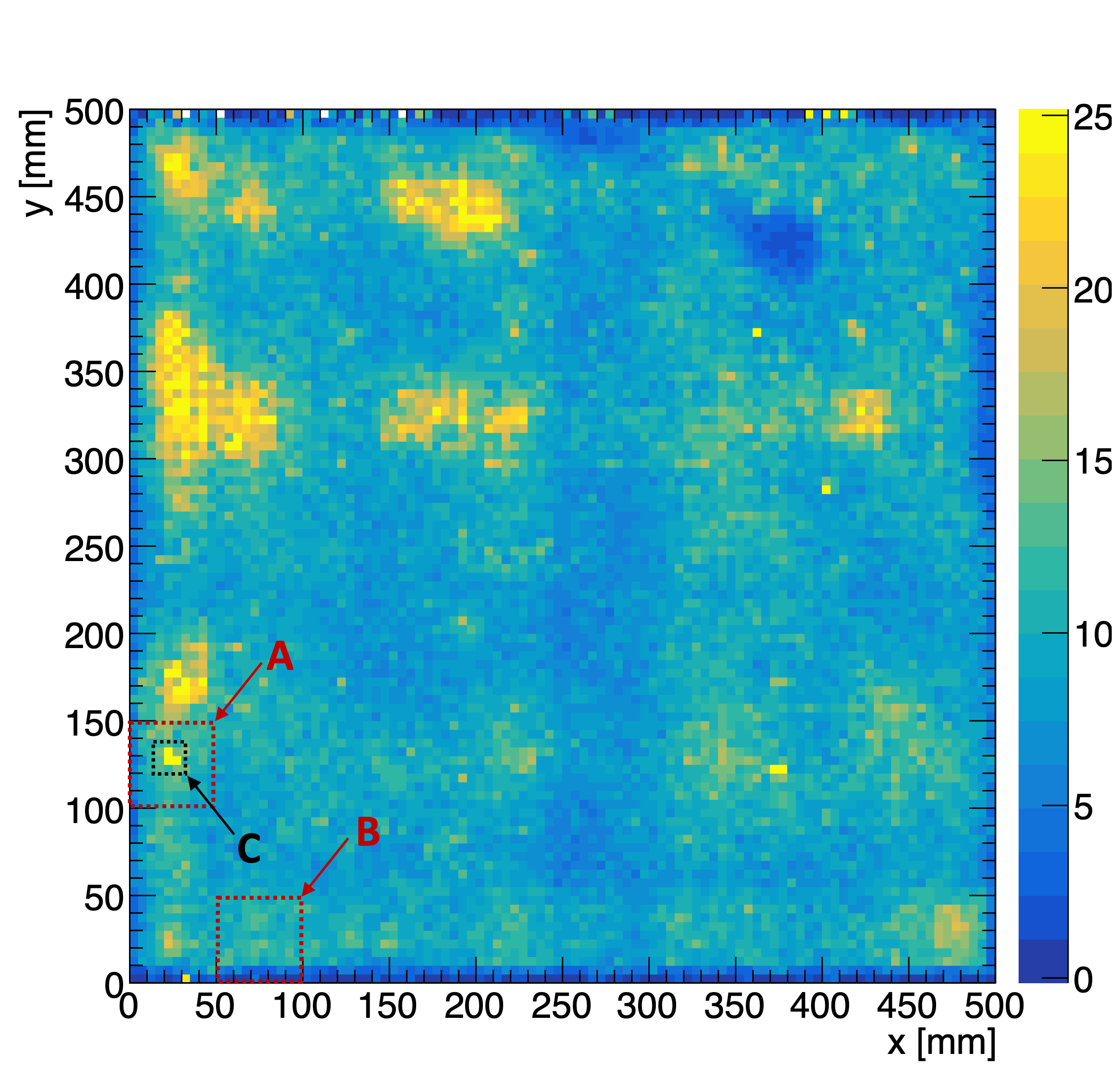}
\caption{\label{fig:charge_map} Average cluster charge map.}
\end{figure}

\begin{figure}
\centering
\begin{subfigure}[b]{0.46\textwidth}
\centering
\includegraphics[width=\textwidth]{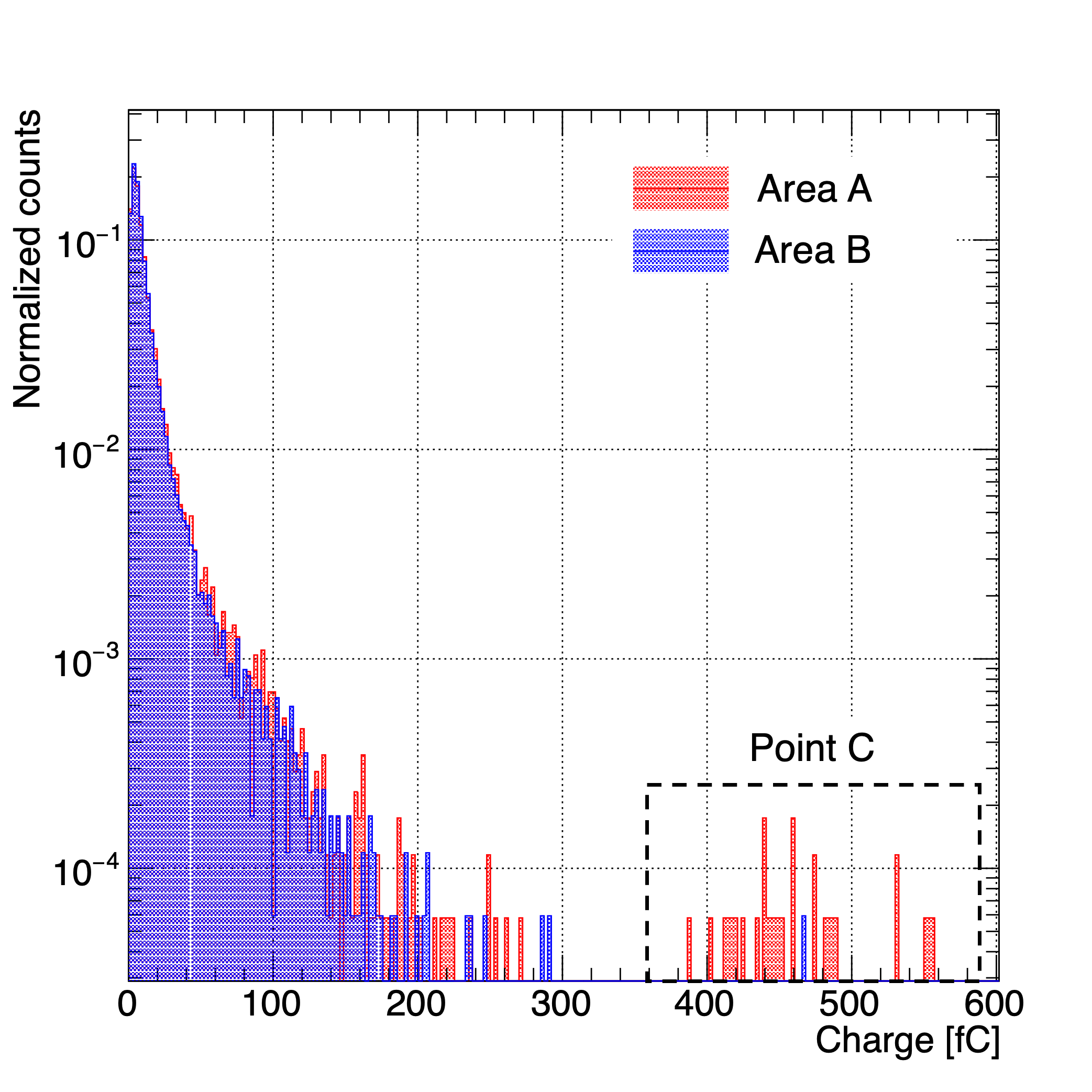}
\caption{}
\label{fig:p-a}
\end{subfigure}
\begin{subfigure}[b]{0.46\textwidth}
\centering
\includegraphics[width=\textwidth]{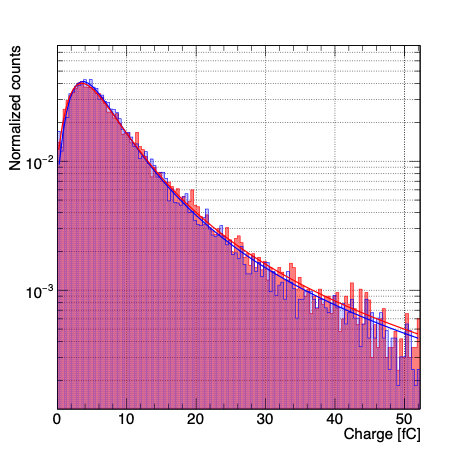}
\caption{}
\label{fig:p-b}
\end{subfigure}


\caption{\label{fig:p} Discharge-like events: a) comparison of charge distribution in the presence of saturated events; b) close-up on the MPV.}
\end{figure}

The average value of the cluster charge distribution measured over areas of $\mathrm{5\times 5 ~mm^2}$ is shown in Figure \ref{fig:charge_map}. The gain non-uniformity pattern resembles the one deduced from the MPV map – Figure \ref{fig:m-a}. In addition, numerous small areas of high average charge are also visible. Figure \ref{fig:p-a} shows the charge distributions measured in areas A and B of Figure \ref{fig:charge_map}. A zoom into the low-charge region (Figure \ref{fig:p-b}) shows a similar response. However, in area B, a second population of high-charge clusters is seen – Figure \ref{fig:p-a}. These muon-induced discharge-like events occur at millimetric hotspots, e.g., point C in Figure \ref{fig:charge_map}. No such events were recorded in the absence of the beam. Following \cite{Jash:2023gpg}, discharges in RPWELL detectors are not associated with current fluctuations or voltage drops, but rather with large signals. Normally \cite{Jash:2022bxy} their magnitude is three orders of magnitude higher than that of avalanche-induced signals, thus, they saturate the APV25 ASIC giving rise to a second charge population distributing around values two orders of magnitude higher than the normal MPV. Point C in Figure \ref{fig:charge_map} is located near a gluing point. Thus, the electrical instability could originate from the assembly imperfections due to the glue residuals penetrating the THWELL holes.

\begin{figure}[htbp]
\centering
\includegraphics[width=.64\textwidth]{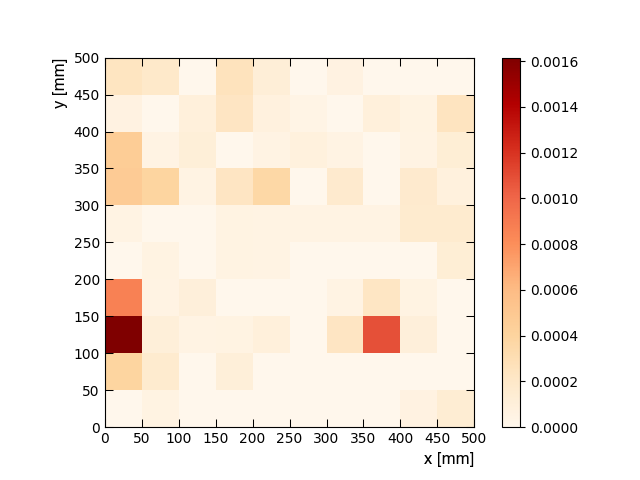}
\caption{\label{fig:q} Discharge rate map.}
\end{figure}

In what follows, a discharge is defined as a cluster with an average strip charge higher than 1500 ADC  counts ($\approx24 ~\mathrm{fC}$). The discharge probability was defined as the ratio between the number of discharges over the total number of events. The discharge rate map is shown in Figure \ref{fig:q}. Discharge probability below $10^{-6}$ is measured across the majority of the detector (upper bounded by the total number of events measured in each region). The discharge probability in regions containing hotspots is of the order of $10^{-3}$.

\section{Summary}
\label{sec:discussion}

 In this work, we introduce a new design and construction technique of $50 \times 50 ~\mathrm{cm^2}$ RPWELL detector. A new prototype was built and studied in $80 ~\mathrm{GeV/c}$ muon beam. Its performance was characterized in terms of detection efficiency, effective gain, and discharge probability.

The average measured detection efficiency is at the level of $96\%$ and the estimated gain is at the level of $7.5\times10^3$. The average gain value is consistent with previously reported values for smaller prototypes and mechanically pressed RPWELL detectors of similar size \cite{Bressler:2015dya, Moleri:2016hgk}. 

Apart from the small region in which the THGEM electrode was detached from the resistive glass, the efficiency and gain uniformity are at the level of 3\% and 22\%, respectively, significantly improved relative to glued prototypes of similar scale \cite{Shaked-Renous:2022kxo, kk}.
The detector stability is improved as well, as indicated by the low discharge probability, lower than $10^{-6}$ over most of the detector area.

The presented design and assembly method can be used for further scaling up the size of RPWELL detectors.

\acknowledgments

This work was supported by Grant No. 3177/19 from the Israeli Science Foundation (ISF), The Pazy Foundation, and the Sir Charles Clore Prize. Special thanks to Martin Kushner Schnur for supporting this research. This research was supported in part by the Nella and Leon Benoziyo Center for High Energy Physics. We thank Givi Sekhniaidze, Eraldo Oliveri, and Yorgos Tsipolitis for their help during the test beam. 

\appendix
\section{Assembly procedure}
\label{Appendix A}

The assembly procedure is conducted in a clean room, and consists of the following steps:

\begin{enumerate}

    \item Preparation of the readout anode: epoxy\footnote{Araldite 2011}/graphite mixture was applied on the surface of the anode with a sponge roller to provide electrical connectivity between the resistive plate and readout strips – Figure \ref{fig:assembly_1}.

    \begin{figure}[htbp]
    \centering
    \includegraphics[width=.7\textwidth]{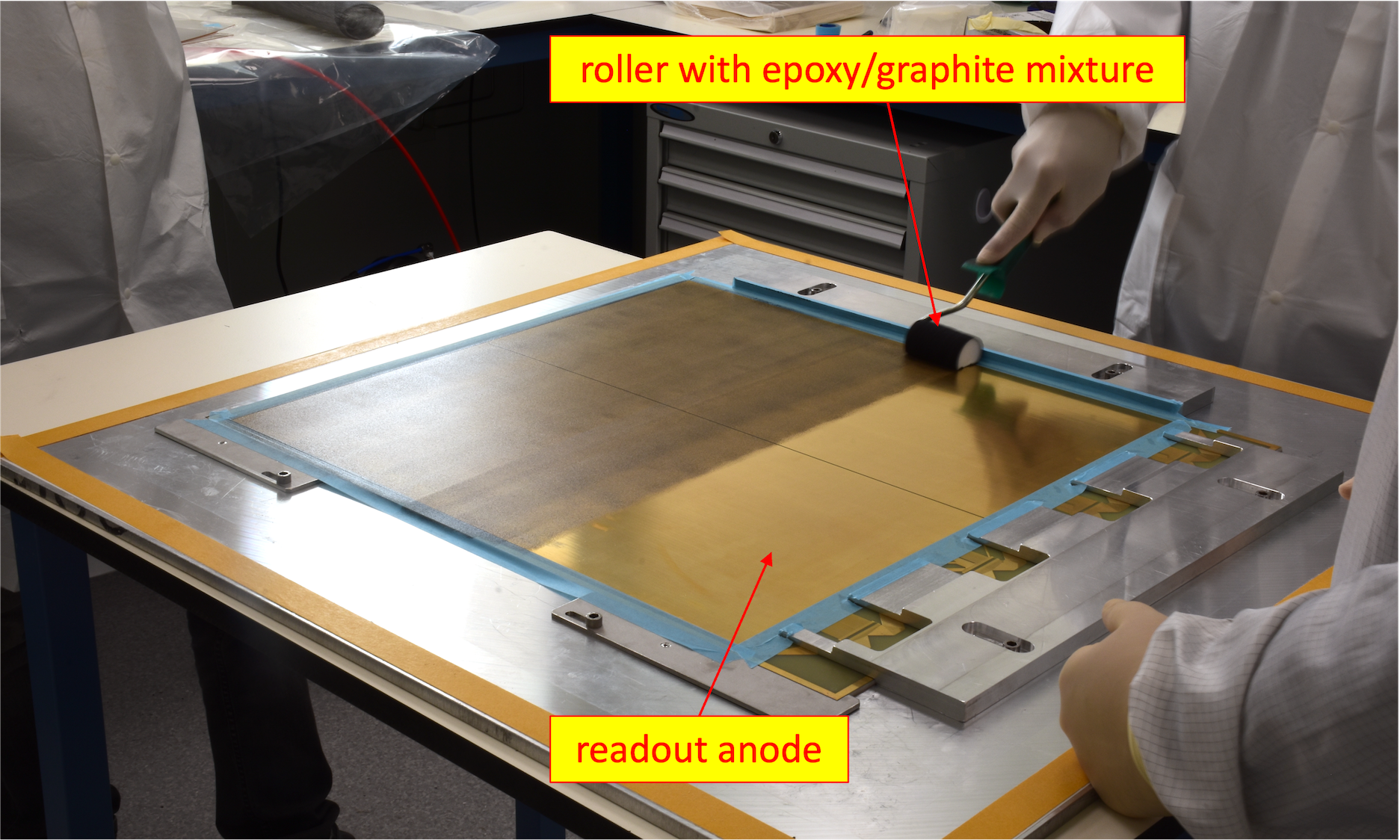}
    \caption{\label{fig:assembly_1} Step 1 of the assembly procedure.}
    \end{figure}     

    \item Resistive plate-to-anode gluing: four glass tiles of $25 \times 25 ~\mathrm{cm^2}$ were placed using precise jigs. The epoxy was cured under pressure enforced with a vacuum bag – Figure \ref{fig:assembly_2}, \ref{fig:assembly_2_1}. 
    
    \begin{figure}[htbp]
    \centering
    \includegraphics[width=.7\textwidth]{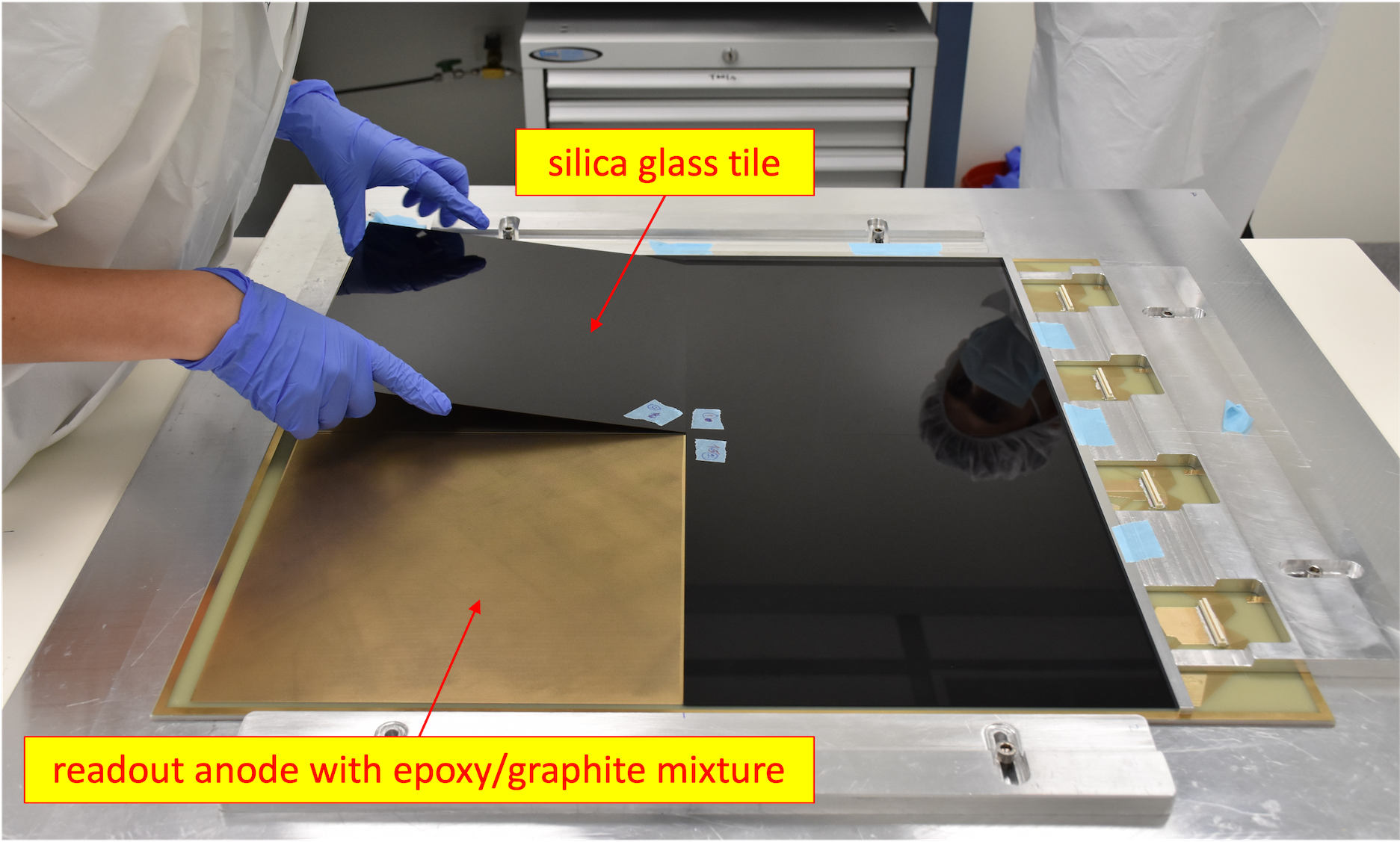}
    \caption{\label{fig:assembly_2} Step 2 of the assembly procedure.}
    \end{figure}

    \begin{figure}[htbp]
    \centering
    \includegraphics[width=.7\textwidth]{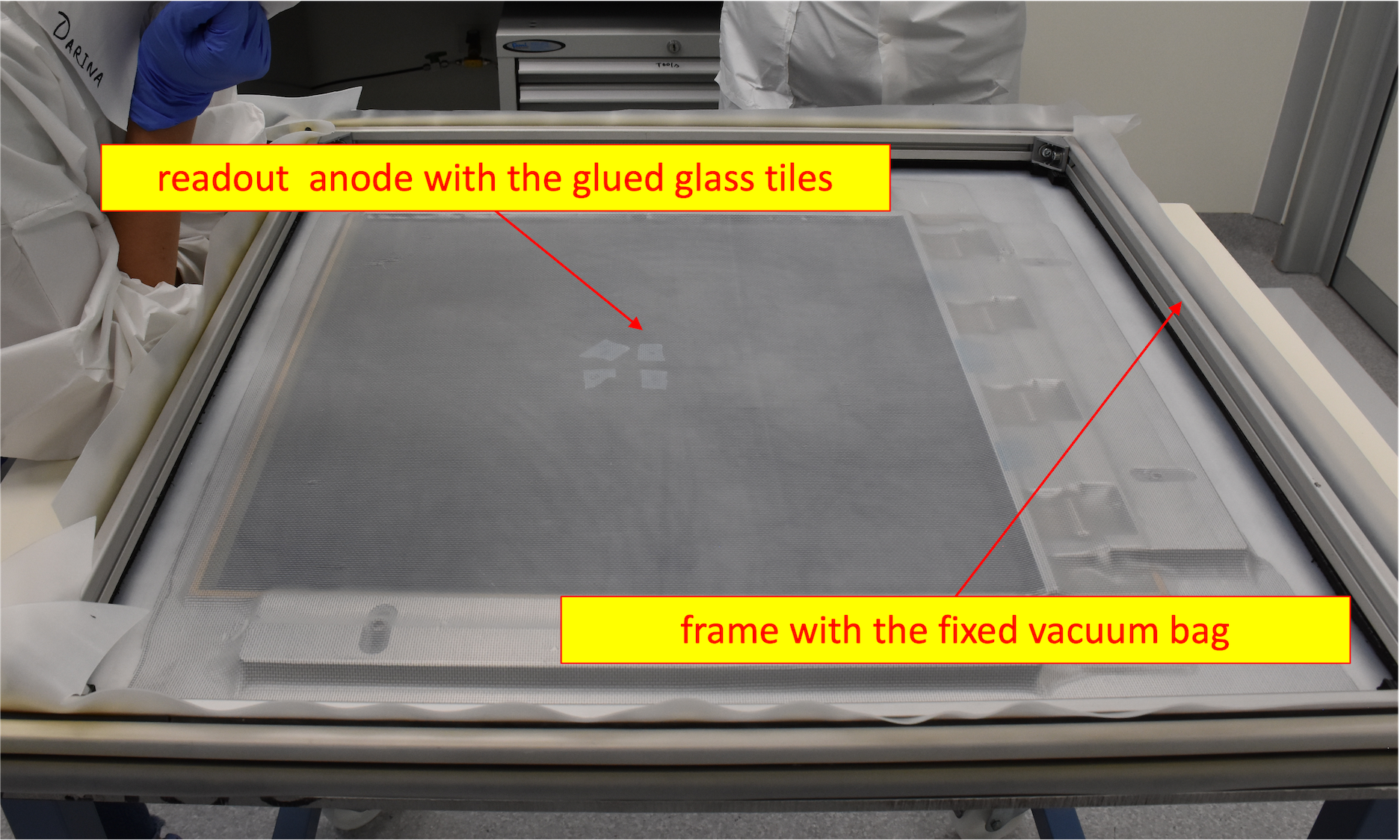}
    \caption{\label{fig:assembly_2_1} Curing under vacuum bag.}
    \end{figure}
    
    \item Coating the edges and intersection of the tiles with insulating resin\footnote{Von Roll Damicoat 2407}  to prevent the occurrence of discharges.
    
    \item Masking THWELL electrode bottom surface: $80 ~\mathrm{\mu m}$ thin adhesive layer was used to cover the  THWELL holes,  leaving 1.2 mm diameter openings for blind spots for glue drops – Figure \ref{fig:assembly_3_1}.
    
    \begin{figure}[htbp]
    \centering
    \includegraphics[width=.7\textwidth]{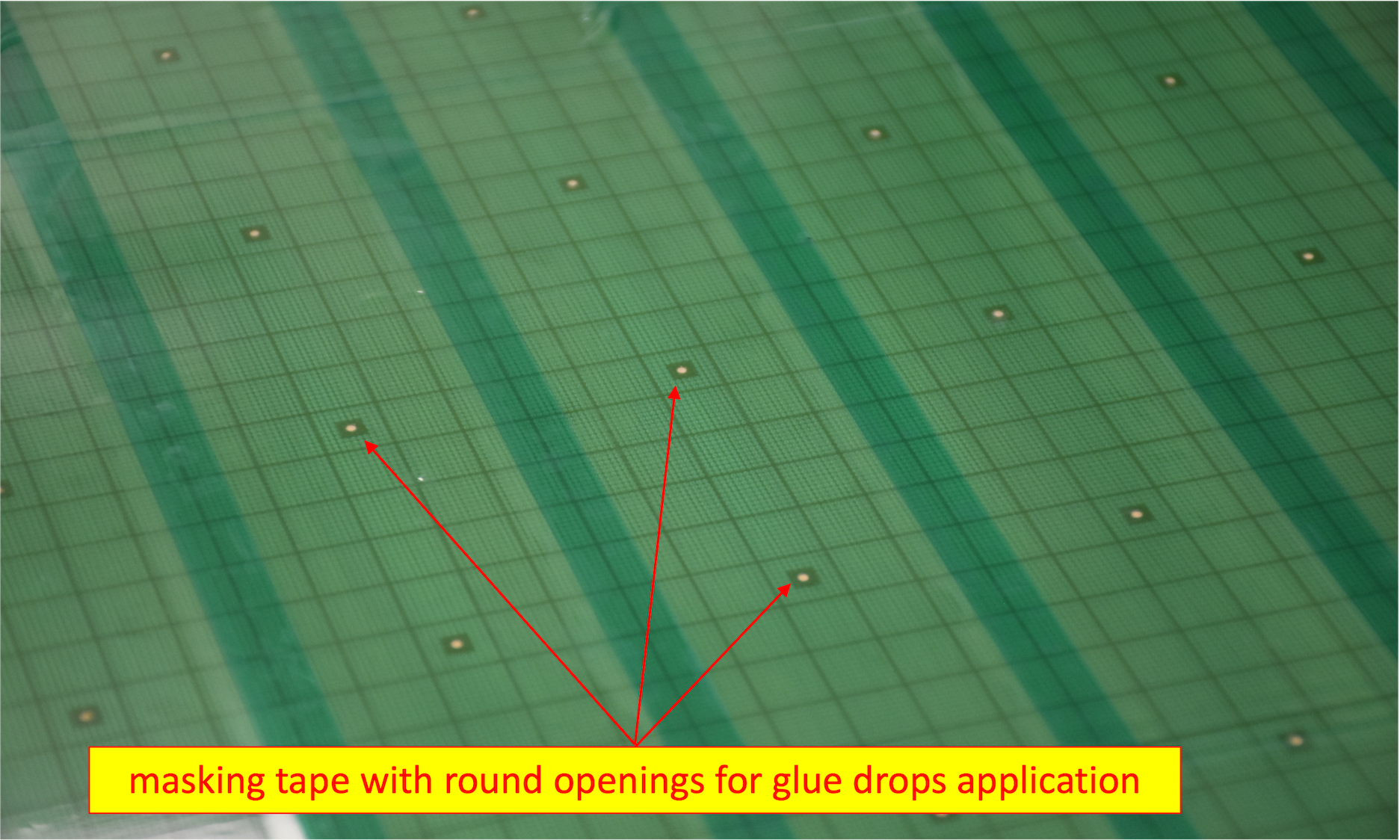}
    \caption{\label{fig:assembly_3_1} Masked bottom surface of the WELL electrode.}
    \end{figure}
    
    \item Applying glue\footnote{Araldite 2011} drops on mask openings using a syringe. Removing excess glue with a plastic spatula and precise metallic jig to assure minimal glue quantity – Figure \ref{fig:assembly_3_2}.
    
    \begin{figure}[htbp]
    \centering
    \includegraphics[width=.45\textwidth]{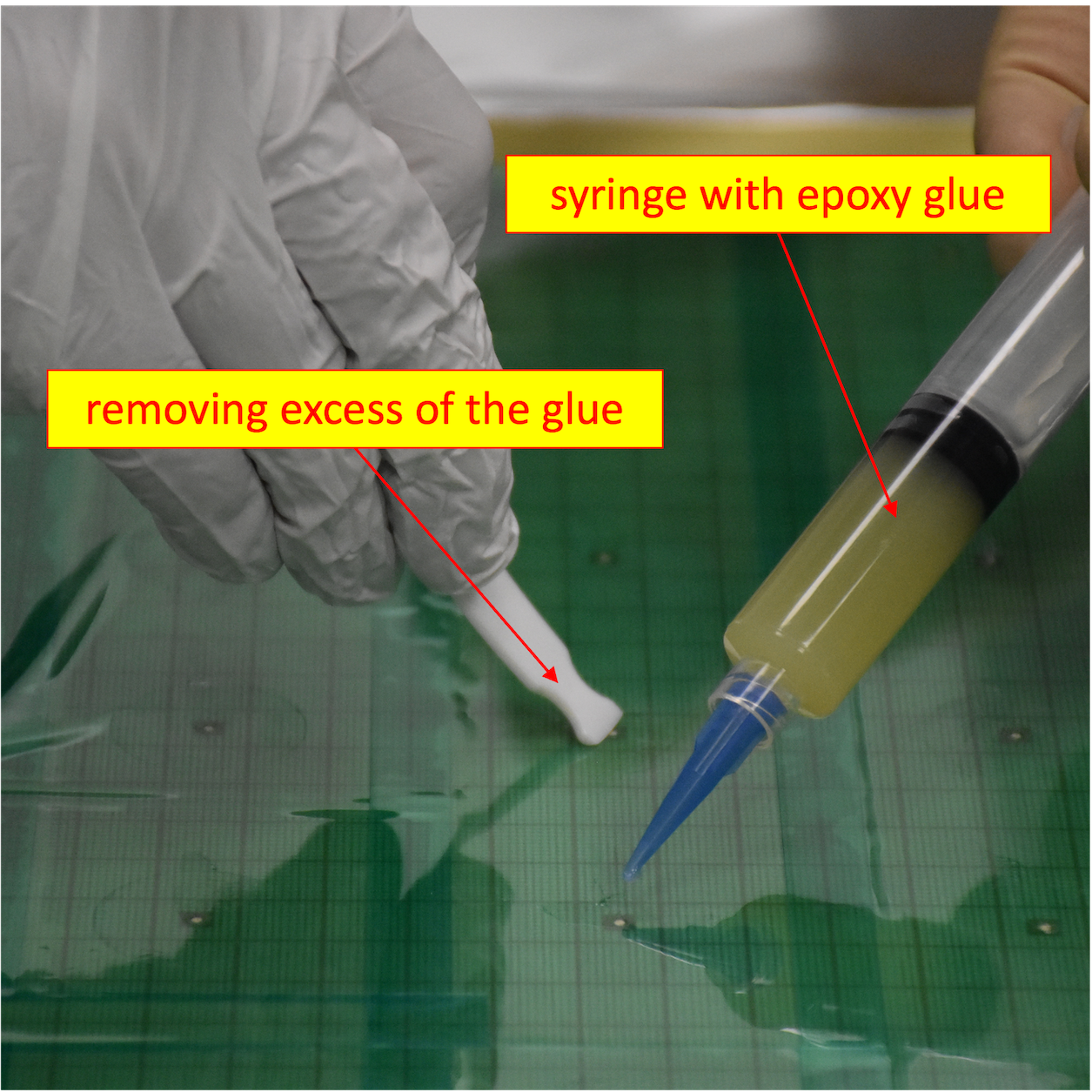}
    \includegraphics[width=.45\textwidth]{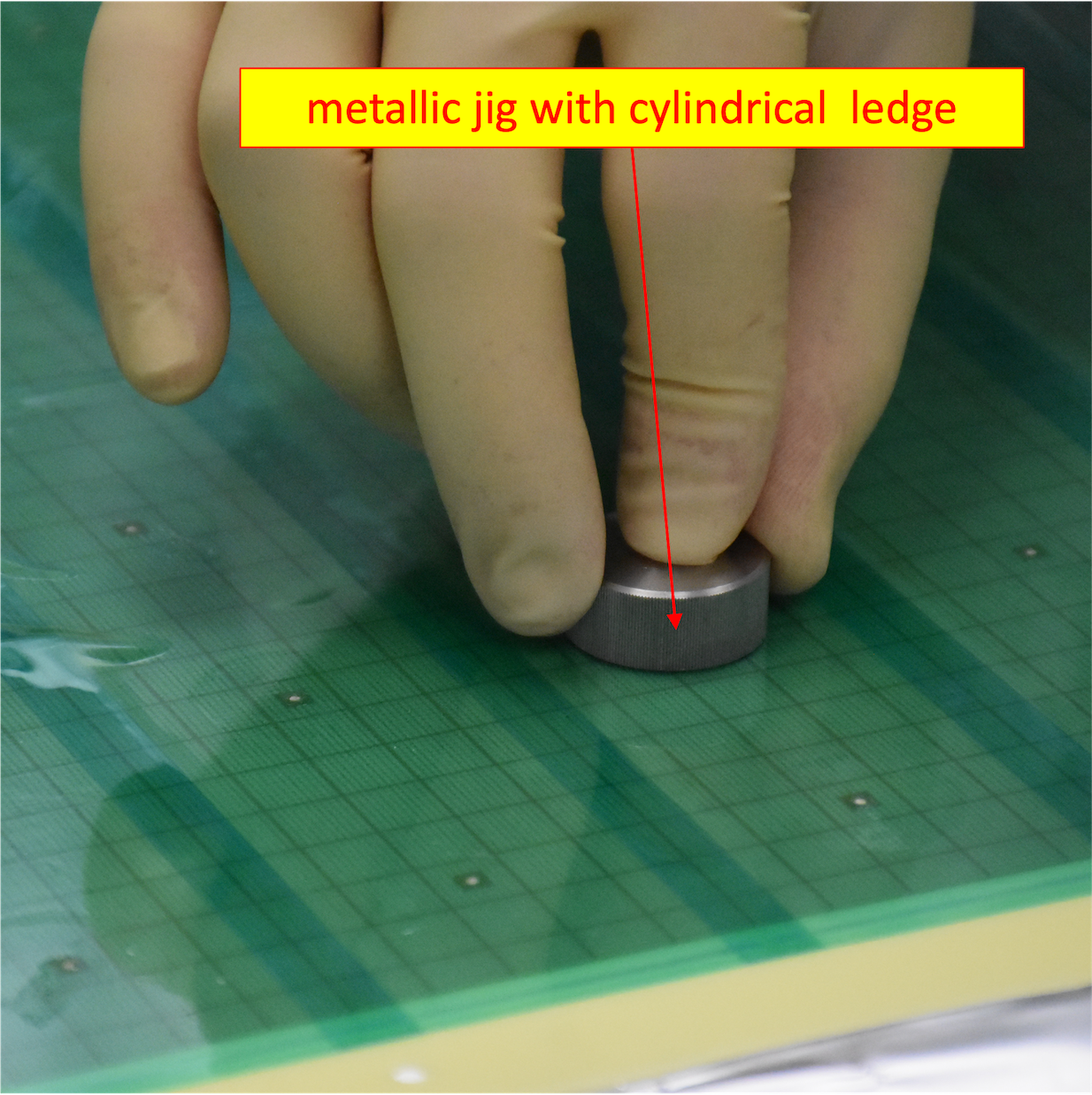}
    \caption{\label{fig:assembly_3_2} Step 5 of the assembly procedure.}
    \end{figure}
    
    \item Gluing the THWELL electrode to the surface of the resistive plate: after unmasking, the electrode with glue drops (Figure \ref{fig:assembly_3_3}) was placed on the surface of the glass using a precise frame – Figure \ref{fig:assembly_4}. 
    
    \begin{figure}[htbp]
    \centering
    \includegraphics[width=.7\textwidth]{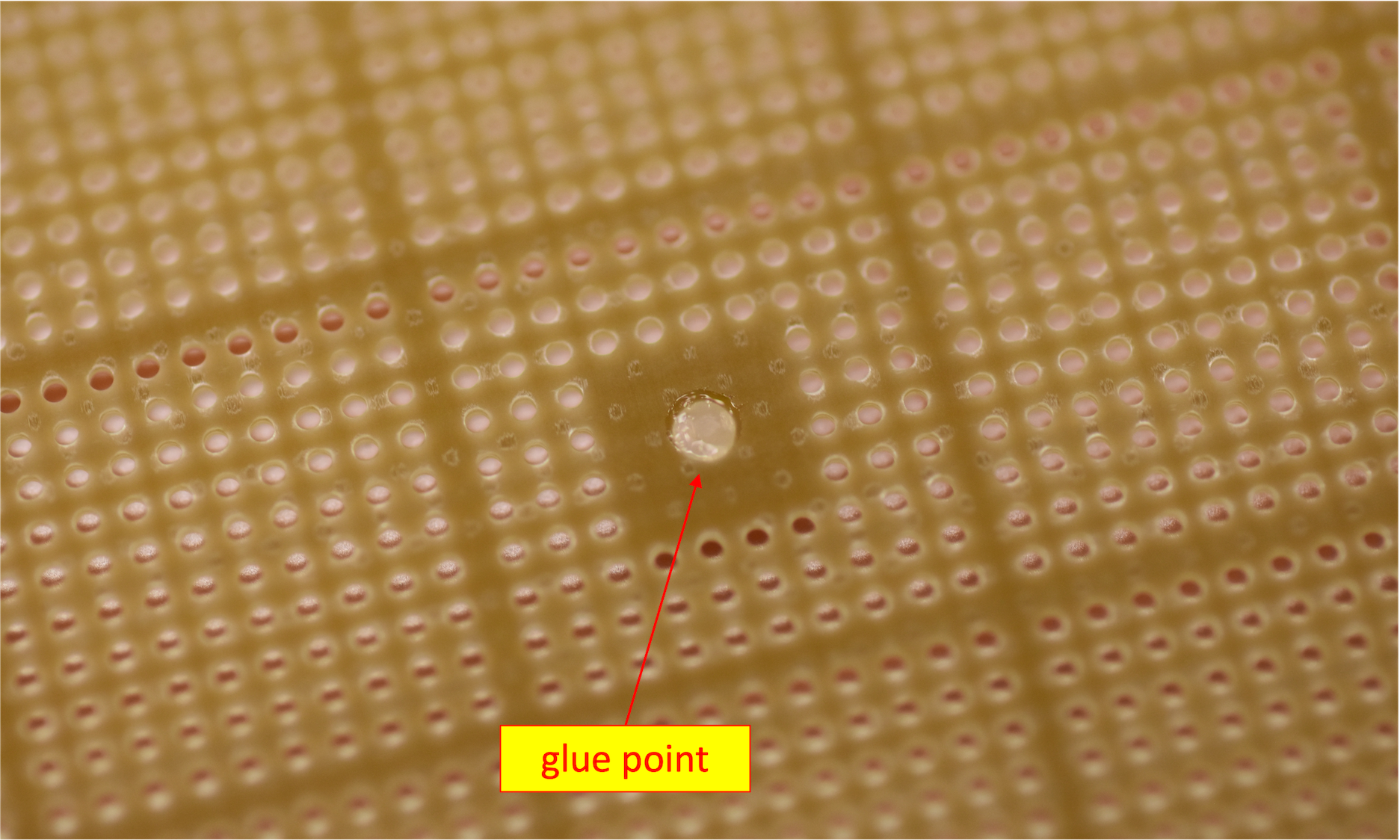}
    \caption{\label{fig:assembly_3_3} Unmasked glue point.}
    \end{figure}
    
    \begin{figure}[htbp]
    \centering
    \includegraphics[width=.7\textwidth]{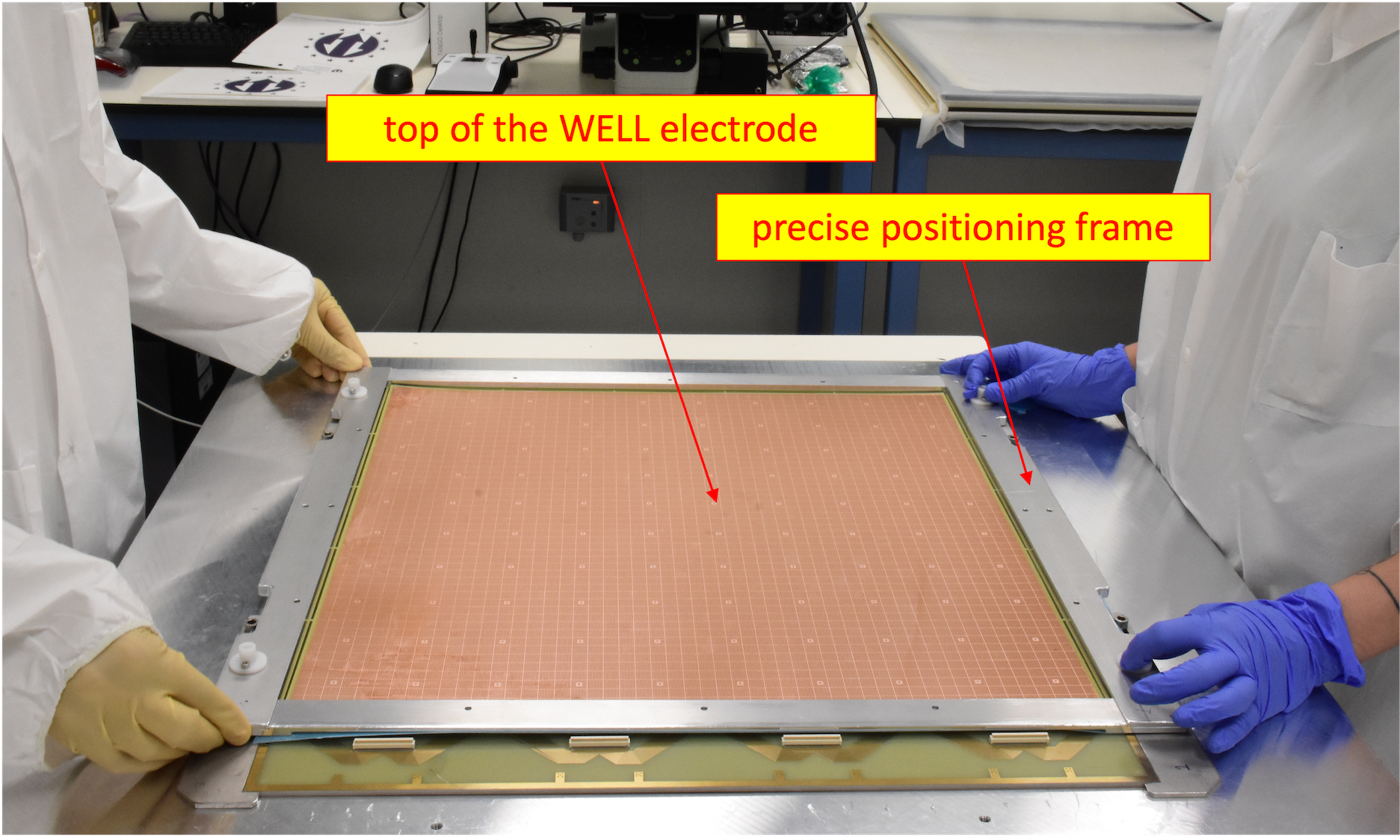}
    \caption{\label{fig:assembly_4} Step 6 of the assembly procedure.}
    \end{figure}
    
    \item Curing the glue points under pressure enforced with a vacuum bag.
    \item Gluing the side frames and closing the chamber with the cathode plate – Figure \ref{fig:assembly_5}.

    \begin{figure}[htbp]
    \centering
    \includegraphics[width=.7\textwidth]{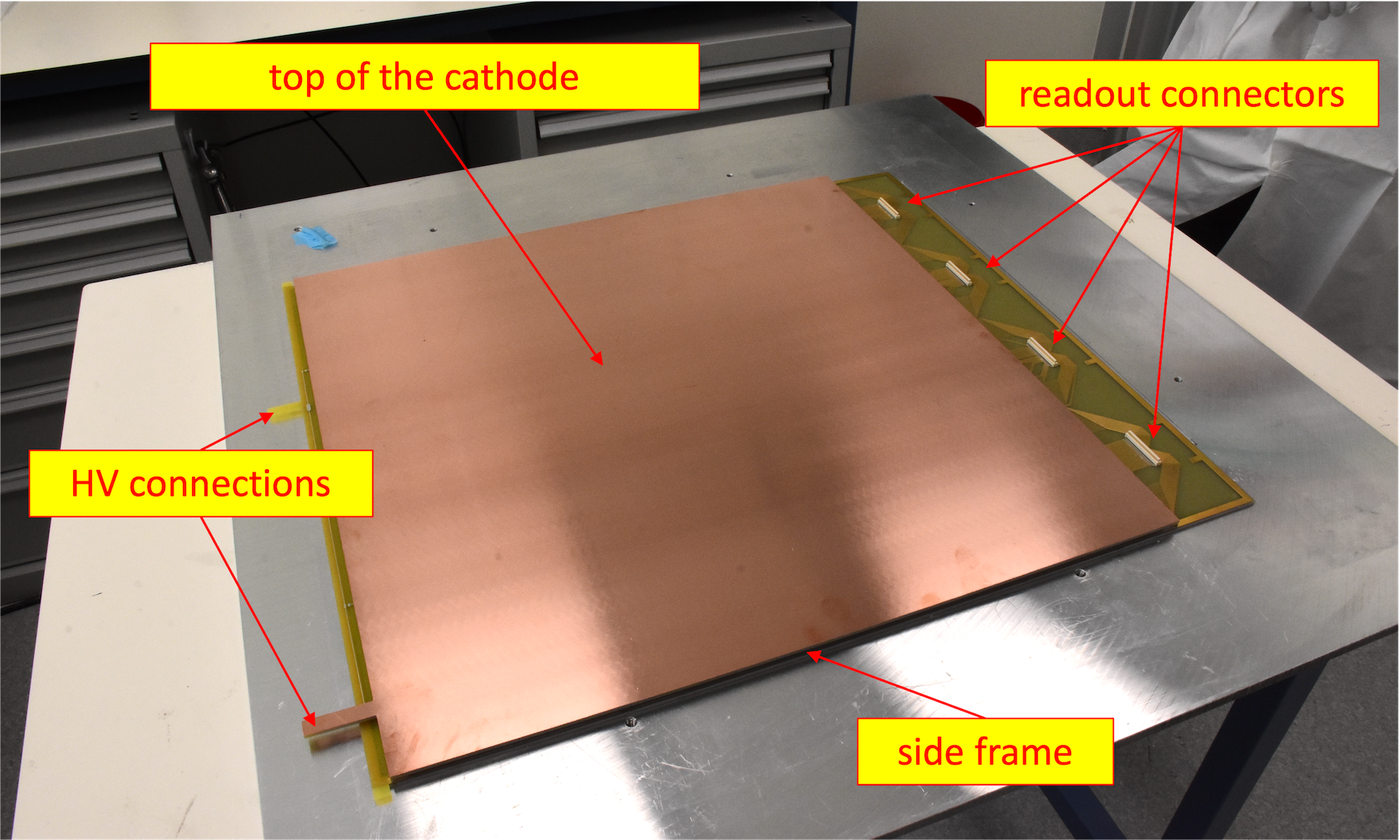}
    \caption{\label{fig:assembly_5} Assembled $50 \times 50 ~\mathrm{cm^2}$ RPWELL chamber.}
    \end{figure}

\end{enumerate}

\end{document}